\documentstyle[multicol,aps,prl]{revtex}

\begin{document}

\draft

\title{
Detection of pairing correlation in the two-dimensional Hubbard model
}

\author{
Kazuhiko Kuroki$^1$, Hideo Aoki$^1$,
Takashi Hotta$^2$, and Yasutami Takada$^2$
}
\address{$^1$Department of Physics, University of Tokyo, Hongo,
Tokyo 113, Japan}
\address{$^2$Institute for Solid State Physics, University of Tokyo,
Roppongi, Tokyo 106, Japan}

\date{\today}

\maketitle

\begin{abstract}

Quantum Monte Carlo method is used to re-examine superconductivity
in the single-band Hubbard model in two dimensions.
Instead of the conventional pairing, we consider
a `correlated pairing',
$\langle \tilde{c}_{i\uparrow} \tilde{c}_{i'\downarrow}
\rangle$
with $\tilde{c}_{i\sigma} \equiv c_{i\sigma}(1-n_{i-\sigma})$,
which is inferred from the $t$-$J$ model, the strong-coupling limit
of the Hubbard model.
The pairing in the $d$-wave channel is found to possess
both a divergence like $1/T$ in the pairing susceptibility and
a growth of the ground-state pairing correlation with sample
size, indicating an off-diagonal long-range order near
(but not exactly at) half-filling.

\end{abstract}

\medskip

\pacs{PACS numbers: 74.20.-z, 71.10.Fd}

\begin{multicols}{2}
\narrowtext


Ever since the discovery of high-$T_c$ cuprate superconductors,
there has been a controversy about its microscopic mechanism.
Following a seminal proposal by Anderson \cite{Anderson} ascribing
it to a purely electronic origin, extensive numerical and analytical
studies have been done to explore a possibility of
superconductivity in the Hubbard model in two dimensions (2D). 
Apart from possible relevance to the high-$T_c$ mechanism, it has
been a long-standing problem to clarify whether the Hubbard model,
a simplest possible model for correlated electrons, can indeed
superconduct \cite{Hirsch}.

A number of studies suggest an effective attraction between
electrons in the model.
For example, two holes are found to be bound in some range of the
on-site repulsion $U$ near half-filling from numerical calculations
for finite systems \cite{Riera,Parola,Dagotto2}.
Quantum Monte Carlo (QMC) results show that the pairing
susceptibility is greater than the susceptibility calculated without
the interaction vertex between quasi-particles
near half-filling \cite{White}, indicating that the quasi-particles
interact attractively.

However, it is not known whether these attractive interactions do
indeed imply a superconducting off-diagonal long-range order (ODLRO).
We can argue against the ODLRO by referring to the fact that
the susceptibility itself for finite $U$ is {\it suppressed} 
from that for $U=0$ \cite{Hirsch3}.
In addition, the susceptibility diverges at most logarithmically
with decreasing temperature $T$, which is exactly the $T$-dependence
for the non-interacting case (inset of Fig.\ \ref{fig1}(a)).
On top of these, the projector Monte Carlo (PMC) calculation for the
ground state \cite{Furukawa} does not show any sign of a growth with
the system size in the equal-time superconducting correlation,
indicating the absence of such an order at least
for $U \leq 4t$, where $t$ is the hopping integral between
nearest-neighbor sites.

Nevertheless, we can still raise
two possibilities that the Hubbard model can indeed exhibit
superconductivity that has eluded detection so far.
One possibility is related to the view that the pairing should be
composed of `quasi-particles' rather than bare particles.
Although this is the point proposed by White {\it et al.}
\onlinecite{White},
they have in fact calculated the conventional BCS
pairing correlation $\langle c_{i\uparrow} c_{i'\downarrow}
c_{j'\downarrow}^\dagger c_{j \uparrow}^\dagger \rangle$
with $c_{i\sigma}^\dagger$ the spin-$\sigma$ electron creation
operator at site $i$ by assuming implicitly that the bare
electrons injected to the system will promptly turn into
appropriately `dressed' ones.
However, as stressed recently by one of the present authors
\cite{Takada}, the dressing may not take place so quickly and
completely in a microscopically small system as usually
encountered in numerical calculations.
In such a situation, we should dress the electrons `by hand'
from the outset to detect superconductivity.

The other, more exotic, possibility is that correlated systems
such as the Hubbard model may possess, in the thermodynamic limit,
a new type of ODLRO that cannot be detected
by $\langle c_{i\uparrow} c_{i'\downarrow}
c_{j'\downarrow}^\dagger c_{j \uparrow}^\dagger \rangle$.
In this case, we should characterize the condensate 
in terms of a certain operator $O$ distinct from
$c_{i\uparrow}c_{i'\downarrow}$ with an ODLRO in the
sense that $\langle O \rangle \neq 0$ breaks the gauge symmetry.

In either possibility, a search for the pertinent correlation
function deserves investigation.
We propose in the present Letter to look at 
$\langle \tilde{c}_{i\uparrow} \tilde{c}_{i'\downarrow}
\tilde{c}_{j'\downarrow}^\dagger \tilde{c}_{j\uparrow}^\dagger
\rangle$ as a candidate
with $\tilde{c}_{i\sigma} \equiv c_{i\sigma}(1-n_{i-\sigma})$
and $n_{i\sigma} \equiv c_{i\sigma}^\dagger c_{i\sigma}$.
In the strong-coupling limit, the Hubbard model can be reduced into
the $t$-$J$ model in which the electrons move with excluded double
occupancies as represented explicitly by the operator
$\tilde{c}_{i\sigma}$.
Thus, in this limit, the conventional pairing correlation
automatically turns precisely into
$\langle \tilde{c}_{i\uparrow} \tilde{c}_{i'\downarrow}
\tilde{c}_{j'\downarrow}^\dagger \tilde{c}_{j\uparrow}^\dagger
\rangle$ \cite{Riera}.
Our proposal amounts that even for 
the Hubbard model with a {\it moderate} $U$,
this can be a proper quantity, for which we have implemented QMC
to obtain the correlation functions for the `correlated pairing'.
The results indeed turn out to exhibit an evidence for superconductivity
in the 2D single-band Hubbard model.

The Hamiltonian ${\cal H}$ for the model is written as
\begin{equation}
{\cal H}=-t\sum_{\langle i,i'\rangle \sigma}
(c_{i\sigma}^\dagger c_{i' \sigma}+{\rm h.c.})+
U\sum_i n_{i\uparrow}n_{i\downarrow}.
\end{equation}
Through a large-scale QMC study, we have evaluated
both the {\it equal-time} correlated pairing
correlation in real space for the {\it ground state},
\begin{equation}
\tilde{p}_\alpha(i,j)=
\langle \widetilde{\Delta}_\alpha^\dagger(i)
\widetilde{\Delta}_\alpha(j)
+\widetilde{\Delta}_\alpha(i)
\widetilde{\Delta}_\alpha^\dagger(j)\rangle ,
\label{eq2}
\end{equation}
and the $k=0$ static {\it correlated pairing susceptibility},
\begin{equation}
\widetilde{S}_\alpha=\frac{1}{N}\sum_{i,j}\int_0^\beta \langle
\widetilde{\Delta}_\alpha(i,0)
\widetilde{\Delta}_\alpha^\dagger(j,\tau)\rangle d\tau,
\end{equation}
as a function of $\beta \equiv 1/T$ in an $N$-site system, where
\begin{equation}
\widetilde{\Delta}_\alpha(i)=\sum_{\delta}f_\alpha(\delta)
(\tilde{c}_{i\uparrow}\tilde{c}_{i+\delta,\downarrow}
-\tilde{c}_{i\downarrow}\tilde{c}_{i+\delta,\uparrow}).
\end{equation}
As for the pairing symmetries we consider 
$d_{x^2-y^2}$ $(\alpha=d)$ with $f_d(x,y)=
\delta_{y,0}(\delta_{x,1}+\delta_{x,-1})-
\delta_{x,0}(\delta_{y,1}+\delta_{y,-1})$
and
extended $s$ $(\alpha=s)$ with $f_s(x,y)=
\delta_{y,0}(\delta_{x,1}+\delta_{x,-1})+
\delta_{x,0}(\delta_{y,1}+\delta_{y,-1})$.

The details of the QMC algorithm are the following.
We have used both the finite-temperature and the
ground-state PMC formalisms.
In both cases, we have employed the discrete
Hubbard-Stratonovich transformation
introduced by Hirsch \cite{Hirsch}.
In the Trotter decomposition, the imaginary time increment
$[\Delta\tau=\tau/$(number of Trotter slices)] is taken to be
either $\leq 0.03$ for finite-temperature susceptibilities or
$\leq 0.07$ in the ground-state pairing correlation.
We have adopted the stabilization algorithm used by several authors
to investigate ground-state and
low-temperature properties \cite{stab}.
Since the negative-sign problem prevents us from obtaining
reliable results for large interactions, we have increased $U$
and/or decreased $T$ up to the point where the ratio of the total
sign to the total number of samples degrades to $\sim 0.3$.

We first show the results for the finite-temperature $d$-wave
pairing susceptibility, $\widetilde{S}_d$, in Fig.\ \ref{fig1}(a).
Calculation is performed for an $8\times 8$ lattice, where the band
filling $n$ is kept to 0.74 
by adjusting the chemical potential 
for each temperature.
For $U=0$ and $U=1$ (in units where $t=1$), $\widetilde{S}_d$
increases only logarithmically with $\beta(=1/T)$.
The logarithmic behavior is similar to that for the
conventional susceptibility $S_d$ shown in the inset.
However, when we increase the repulsion to $U=3$,
$\widetilde{S}_d$ starts to increase more rapidly and can no longer
be fitted to a logarithmic form within the error bars.
Strikingly, the behavior fits rather well to a linear ($\propto 1/T$)
dependence for $\beta \geq 5$.
The $1/T$ behavior is indicative of a long-range order,
as exemplified by the $1/T$ divergence of the
staggered magnetic susceptibility of the single-band Hubbard model
at half-filling \cite{Hirsch,sdw}.
By contrast, $\widetilde{S}_s$ in Fig.\ \ref{fig1}(b) is strongly
suppressed when $U$ is switched on, and barely increases
with $\beta$.

When the band filling is decreased to $n\sim 0.6$, $\widetilde{S}_d$ shows
much weaker dependence on either $\beta$ or $U$ 
(Fig.\ \ref{fig2}(a)).  
There the $\beta$ dependence is so small that we cannot 
even decide whether 
it is linear or logarithmic within the error bars.  
If we increase the band filling to the half-filling ($n=1$), 
on the other hand, 
$\widetilde{S}_d$ is suppressed when $U$ is switched on,
where the susceptibility seems to assume a logarithmic $\beta$-dependence 
as indicated in Fig.\ \ref{fig2}(b).
These results suggest that the strong divergence of
$\widetilde{S}_d$ with $\beta$ occurs only in a {\it limited} region of
$n$ near half-filling.

Now we turn to the equal-time pairing correlation,
$\tilde{p}_\alpha(i,j)$ (Eq.(\ref{eq2})), in the ground state.
We represent its real-space behavior by the quantity
$\widetilde{P}_\alpha (R)$, defined by the sum of
$\tilde{p}_\alpha({\bf r},{\bf r}+ \Delta {\bf r})$ on a square
at a distance $|\Delta r_x|=R$ or $|\Delta r_y|=R$.
This sum is intended to reduce statistical errors.
In Fig.\ \ref{fig3}, we show the ratio of $\widetilde{P}_\alpha(R)$
for $U=3$ to that for $U=0$ for both $d_{x^2-y^2}$ and
extended $s$ channels for a $10\times 10$ system with $n=0.82$.
For the $d$-wave ($\tilde{P}_d(R)$), the pairing correlation at large 
distances is indeed enhanced by $U$,
while the correlation is strongly suppressed for the extended $s$.  
There we should note that the results at $R=$ 
half the sample size suffer from an effect of periodic
boundary condition.

An increase of the correlation function with $U$ does not necessarily
signify the occurrence of ODLRO as stressed by various authors.
An appropriate quantity for this purpose is the $k=0$ Fourier
component, {\it i.e.}, $\sum_j\tilde{p}_\alpha(0,j)
=\sum_R \widetilde{P}_\alpha(R)$ of the real-space correlation, 
and its system-size dependence in particular.
This quantity should grow with the system size when a long-range
order sets in.

The QMC calculation is done for this quantity in two different
sizes: 
a $10\times 10 $ lattice with 74 electrons $(n=0.740)$, and
a $14\times 14$ lattice with 146 electrons $(n=0.745)$.
The choice of these sizes is made due partly to the fact that the
band fillings are almost the same, but we also have paid special
attention to the fillings satisfying the closed-shell condition,
{\it i.e.}, band fillings for which the non-interacting Fermi sea
is non-degenerate.
(The pairing correlation at open-shell fillings is inherently smaller, 
so that a mixture of results for close- and open-shell fillings 
would obscure the conclusion \cite{open}.)
The closed-shell condition also reduces greatly the difficulty
arising from the negative-sign problem in the PMC calculation.
The above two fillings are the closest closed-shell ones
within the system sizes tractable in the QMC.
Even for such close fillings, $\widetilde{P}_d(R)$
at very short distances ($R=0$ or $1$) has a considerable filling
dependence due primarily to the filling dependence of density
(diagonal) correlations contained inherently in the pairing
correlation.
In order to focus on the longer-range (off-diagonal) part, we
have evaluated $\widetilde{P}_d$ by the sum of $\widetilde{P}_d(R)$
for $R \geq 2$.
The values for $\widetilde{P}_d$'s thus obtained for the two system
sizes almost coincide (with the difference being less than 0.0008)
for $U=0$ \cite{comment}.
The same is true as long as $U$ is smaller than 1.5.

When $U$ is further increased, the long-range part of the
correlation grows with $U$, where $\widetilde{P}_d$ does concomitantly
start to increase with the sample size as seen in Fig.\ \ref{fig4}.  
To be more precise, the amount of this increase exceeds 
the error bar for $U$ larger than about 2.
Although there still remains, to be rigorous, 
a possibility that this 
size-dependence contains an effect of the slight discrepancy
in the filling, the result for the correlation may be taken as an indication
toward the ODLRO, which reinforces the conclusion drawn from
the $T$- dependence of the susceptibility.

We expect that the long-range behavior of $\widetilde{P}_d$ and
$\widetilde{S}_d$ continues to grow for larger $U$, in which the present 
correlated pairing correlation will eventually 
tend to the conventional one.
We can then envisage a consistent picture that encompasses 
a recent result\cite{Kuroki} for 
the Mott-Hubbard regime of Emery's three-band Hubbard model\cite{Emery}, 
where an indication of a long-range order in the conventional $d$-wave
pairing is detected for $U_d$ comparable to the 
one-electron (anti-bonding $d$-$p$) bandwidth.

As for the nature of superconductivity for moderate $U$, further
investigation is necessary in the following two respects.
First, we cannot at the present stage decide whether the correlated
pairing either represents just a device for detection of
superconductivity in small systems or indicates a new ODLRO.  
Second, extensive studies for the $t$-$J$ model in 1D\cite{Ogata} 
and 2D\cite{Dagotto} suggest
that superconductivity appears for a large enough superexchange
interaction $J/t$ with apparently the same form of pairing as ours.
However, these studies focus on a parameter region distinct from
the present study.  
Remember that if the $t$-$J$ model is regarded as an effective
Hamiltonian for the Hubbard model for $U\rightarrow \infty$, 
we end up with a perturbatively
small $J/t$.
Thus whether the mechanism for
superconductivity in the $t$-$J$ model with a large $J/t$
is similar to the present case remains to be another future problem.

In summary, we have shown from the results of the pairing
susceptibility and the correlation function obtained
in the QMC method that the `correlated pairing'
with a $d_{x^2-y^2}$ symmetry can become long-ranged in the 2D
Hubbard model with moderate $U$ near half-filling.

Numerical calculations were done on HITAC S3800/280
at the Computer Center of the University of Tokyo,
and FACOM VPP 500/40 at the Supercomputer Center,
Institute for Solid State Physics, University of Tokyo.
For the former facility we thank Prof. Y. Kanada for a
support in `Project for Vectoralized Super Computing'.
This work was also supported in part by Grant-in-Aids for Scientific
Research from the Ministry of Education, Science, Sports and Culture
of Japan.


\begin{figure}
\caption{
(a) The correlated $d$-wave pairing susceptibility,
$\widetilde{S}_d$, for a fixed band filling $n\simeq 0.74$
is plotted against $\beta(\equiv 1/T)$ for $U=3$
$(\bigcirc)$ or $U=1$ $(\triangle)$,
while the solid curve represents the non-interacting case.
The dashed curve is a logarithmic least-squares fit for $U=3$
(for $4\leq\beta\leq6)$, the dash-dotted line a linear fit
for $U=3$ (for $5\leq\beta\leq10)$, and the dotted curve
a logarithmic fit for $U=1$ $(4\leq\beta\leq10)$.
The inset depicts the conventional $d$-wave pairing susceptibility,
$S_d$, for comparison, where $\bigcirc$ represents
the result for $U=3$ with a logarithmic fit for $4\leq\beta\leq10$,
while the solid curve indicates the non-interacting case.
(b) The correlated extended $s$-wave pairing susceptibility,
$\widetilde{S}_s$, for $U=3$ $(\bigcirc)$ with a logarithmic fit for
$4\leq\beta\leq10$ (dashed line), or for $U=0$ (solid line).
}
\label{fig1}
\end{figure}

\begin{figure}
\caption{
(a) $\widetilde{S}_d$ as a function of $\beta$ for a fixed $n\simeq 0.6$
with $U=3$ $(\bigcirc)$ and the non-interacting case (solid curve).
(b) A similar plot for $n=1$, where the dashed curve is a logarithmic
fit for $U=3$.
}
\label{fig2}
\end{figure}

\begin{figure}
\caption{
The ratio of the correlation function $\widetilde{P}_\alpha(R)$
for $U=3$ to that for $U=0$ is plotted against $R$ for the $d$
$(\bigcirc)$ or extended-$s$ $(\triangle)$ pairing.
The system size is $10\times 10$ with 82 electrons ($n=0.82$).
}
\label{fig3}
\end{figure}

\begin{figure}
\caption{
$\widetilde{P}_d$ is plotted against $U$.  
Number of electrons and system size are 
146 electrons$/ 14 \times 14$ $(\bigcirc, n=0.745)$ and 
74 electrons$/10\times 10$ $(\Box, n=0.740)$.
}
\label{fig4}
\end{figure}

\end{multicols}

\begin{references}
\bibitem{sdw} Away from the half filling, a QMC result\cite{Furukawa}
shows that the SDW correlation is short-ranged.
\bibitem{comment} If we include the $R=0,1$ components,
$\widetilde{P}_d$ for two sizes would differ by 0.013 for $U=0$,
which would obscure the size dependence of the long-range
part of the correlation.
\end{references}
\end{document}